\documentclass[journal]{IEEEtran}

\newcommand{\etal}{{\it et~al.}}
\newcommand{\ie}{{\it i.e., }}

\usepackage{amsmath, scalerel}

\usepackage[linesnumbered,ruled]{algorithm2e}
\usepackage{graphicx}
\graphicspath{ {images/} }
\usepackage{pgfplots, lipsum}
\usepackage{textcomp}
\pgfplotsset{compat = newest}

\pgfplotsset{width=10cm,compat=1.9}
\usepgfplotslibrary{external}
\usepackage{tikz}
\usetikzlibrary{arrows}
\usepackage{caption}
\usepackage{siunitx}
\usepackage[font={small}]{caption}
\usepackage{amsmath}
\usepackage[export]{adjustbox}
\usepackage{lipsum}
\usepackage{subcaption}

\usepackage{xcolor}
\usepackage{pgfplots}
\usepackage{tikz}
\usepackage{pgfkeys}
\usepackage{url}
\usepackage{pgfplotstable}
\usepackage{subcaption}
\usepackage{filecontents}
\usepackage{multirow}
\usepackage{pgf}

\makeatletter
\def\BState{\State\hskip-\ALG@thistlm}

\makeatother

\usepackage{cite}

\ifCLASSINFOpdf
  
\else
  
\fi

\begin{document}

\title{ Cost-Efficient Storage for On-Demand Video Streaming on Cloud
}

\author{\IEEEauthorblockN{Mahmoud Darwich$^{1}$, Yasser Ismail$^{2}$}, Talal Darwich$^{3}$, Magdy Bayoumi$^{4}$\\ 
\IEEEauthorblockA{ $^{1}$Department of Mathematical and Digital Sciences, 
Bloomsburg University of Pennsylvania, PA 17815\\
$^{2}$Electrical Engineering Department, 
Southern University and A\&M College, Baton Rouge LA 70807\\
$^{3}$Microchip Technology Inc., San Jose, CA 95134\\
$^{4}$Department of Electrical and Computer Engineering,
University of Louisiana at Lafayette, LA 70504 \\
Email: mdarwich@bloomu.edu, yasser\_ismail@subr.edu, talal.darwich@microchip.com, magdy.bayoumi@louisiana.edu}
}

\maketitle

\begin{abstract}

Video stream is converted to several formats to support the user's device, this conversion process is called video transcoding, which imposes high storage and powerful resources. With emerging of cloud technology, video stream companies adopted to process video on the cloud. Generally, many formats of the same video are made (pre-transcoded) and streamed to the adequate user's device. However, pre-transcoding demands huge storage space and incurs a high-cost to the video stream companies. More importantly, the pre-transcoding of video streams could be hierarchy carried out through different storage types in the cloud. To minimize the storage cost, in this paper, we propose a method to store video streams in the hierarchical storage of the cloud. Particularly, we develop a method to decide which video stream should be pre-transcoded in its suitable cloud storage to minimize the overall cost. Experimental simulation and results show the effectiveness of our approach, specifically, when the percentage of frequently accessed videos is high in repositories, the proposed approach minimizes the overall cost by up to 40\%.
\end{abstract}


\IEEEpeerreviewmaketitle

\section{Introduction }

Video streaming has become widely used in electronic displaying devices-based applications. Due to the huge number of videos that are streamed on a variety of devices such as large screen TVs, desktops, tablets, and smart-phones. Video streaming is the main source of Internet traffic in the United States. It consumes up to 77\% of the Internet Bandwidth in the United States \cite{report}. Additionally, video streaming is expected to consume up to 85\% of Internet traffic by 2021 \cite{forecast}.
Based on the characteristics of the end-device of the video streaming; i.e. the allowed bit-rate, resolution, and network bandwidth; the Video contents have to be transcoded to match the characteristics of the end-device \cite{ahmad}. Video On-Demand (VOD) such as YouTube or Netflix and live-streaming such as Livestream are examples of video content.
Video transcoding is an exhaustively time and computation consuming process. Cloud computing services have been used by the Video Stream Providers (VSP) to greatly decrease the overall computations of the transcoding process \cite{xiangbo}. The VSPs perform the transcoding operation offline on the VOD to guarantee high-speed video streaming operations. During the offline transcoding operation, multiple formats of the video stream are stored on a cloud. Based on the specifications of the end-device of the viewer, the proper stored format will be selected from the cloud. The process of storing multiple formats of the video stream is called pre- transcoding. As a practical example of the pre-transcoding operation, Netflix pre- transcodes and stores approximately 70  different formats of each video on their cloud \cite{netflix}. As a result, there will be an extra overhead cost for the VSPs \cite{livlsc,li1}.

The distribution of accessing the pre-transcoded videos is a long tail distribution \cite{sharma}. It means many pre-transcoded videos are accessed while a small number of such videos are rarely accessed. This encourages researchers to reduce the overall pre-transcoding cost by transcoding rarely-accessed videos as on-demand videos \cite{xiangbo,li1}. This will allow one or a few formats of a video to be stored and transcoding is executed when accessing a video format that is not already pre-transcoded. We term the rarely-accessed videos (lazy transcoding of videos) as re-transcoding and storing of videos as pre-transcoding.

The  cost of cloud VMs  is higher than the cloud storage cost \cite{amazon}. This is because the computational cost is calculated per hour in the cloud. This indicates that the re-transcoding operation is cost efficient to VSPs when applying it to the rarely accessed videos. By contrast, if pre-transcoding is executed on frequently accessed videos (FAVs), it results in a very high cost because it charges VSPs  each time the video is transcoded. This is why the pre-transcoding approach is alternatively applied to frequently accessed videos. 

The research problem of the proposed work is how to decide where the video stream should be stored in the cloud storage. Therefore, a method that performs clustering on the frequently accessed video streams is proposed in this paper to tackle this issue.

The main contributions of this paper can be summarized as follows: (1) Proposing a method to reduce the incurred cost of using cloud services through clustering the frequently accessed video streams in the repository; and (2) Analyzing the effectiveness of the proposed method when changing the number of frequently accessed video streams in a repository . This proposed work differs from the previous work \cite{darwich2016} in that we design an approach that stores video streams efficiently and thus it decreases further the cost of cloud services.

The paper is organized as follows: section II reveals the related work. The clustering method will be explained in section III. Experiment setup and results are explained in sections IV and V, respectively. Conclusion and future work are presented in sections VI respectively.

\section{Related Work}

Kim \etal \cite{kimmultimedia2019} proposed a scheme to transcode multimedia video streams resources using intra-cloud and the parallel computing framework. Their scheme provided improved tasks assignment and high-speed video transcoding.

Darwich \etal \cite{darwichhotness} proposed algorithms to reduce the incurred cost paid by VSPs when using cloud services. Particularly, they developed a method that measures how much frequently the video stream is accessed and accordingly, their approach decides whether the video stream to be stored in the repository or transcodes it upon request.

Gao \etal \cite{gao} proposed an approach to transcode video partially using the cloud. Their proposed method store the frequently accessed segments of video and which are located in the beginning, while they drop the remaining segments and transcode them upon request. Their method reduced 30\% of the cost compared to storing all segments of the video.

Zhao \etal \cite{zhao} developed a method that reduces the operational cost of video streaming on the cloud. Particularly, their approach trades off between the transcoding and storing the video, they implemented it by using the weight graph of the video transcoding. They used the transcoding relationships between video and their popularity to decide the video versions that should be kept and stored or dropped and re-transcoded upon request.

 Jokhio \etal \cite{jokhio} proposed an approach that estimates the costs of storing and transcoding a video using cloud resources. Besides, their approach utilizes the popularity of each transcoded video to come up with a decision about the time frame for storing or re-transcoding it. Their results show the efficiency of the method by reducing the cost significantly.

\section{Proposed Clustering Method}
\subsection{ Structure of Video Stream}
\begin{figure}
 \hfill\includegraphics[width=7.47cm]{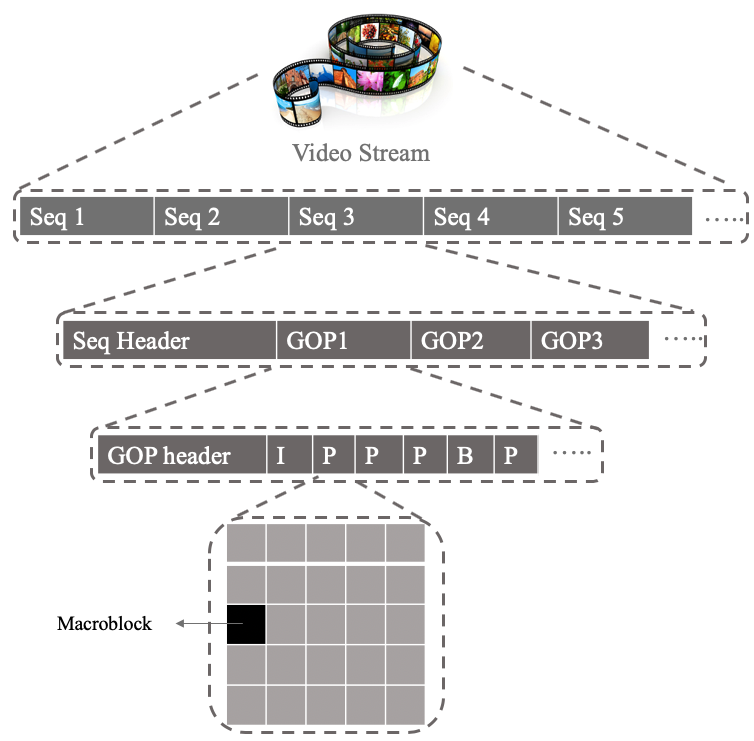}
 \caption{Structure of a video stream}
 \label {video-stream-structure}
\end{figure}

A Video is composed of many sequences as illustrated in Fig.\ref{video-stream-structure}. The sequence in the video stream is formed by Group Of Pictures (GOPs). The structure of a sequence and GOP is started with sequence header and GOP header respectively. The headers include meta-data about sequence and GOP. Different types of frames are contained in a GOP (\ie I (intra), P (predicted), and B (bi-directional) frames). Further, each frame is composed of tiny slices called macroblocks (MB) \cite{jokhio1}. 
 
 The operation of video transcoding is carried out at the GOPs level because they can be processed independently \cite{jokhio1}. Thus in this research, we considered the transcoding process at the GOPs level.
 
\subsection{Video Streaming Using Cloud}
The cloud services are available in an on-demand way. That means the users are charged in a pay-as-you-go way. Video streaming on the cloud requires the following services:
\begin{itemize}
\item Computational Services: The transcoding operations of videos are achieved using Virtual Machines (VM) and the charge is an hourly basis
\item Storage Services: Cloud providers offer different storages for users and the charge is a monthly basis.
\end{itemize}
Amazon Web Services is a well know company for cloud services, it offers cloud services with affordable price and high reliability. Although we consider AWS services and models in our study. This is research could be applied to any cloud services.

Amazon offers different types of storages. \texttt{S3 Standard} Storage, S3 Standard-Infrequent Access (\texttt{S3 Standard-IA}) Storage, S3 One Zone-Infrequent Access (\texttt{S3 One Zone-IA}) Storage, and  \texttt{S3 Glacier }Storage. The Amazon storage services are rated for each Gigabyte of stored data in a month.
These storage services are based on different bandwidth accesses at different rates.

\subsection{Algorithm }

The proposed algorithm is an improved version of the previous work \cite{darwich2016}. Its purpose is to reduce the cost of video streaming on the cloud by applying clustering on the frequent accessed video/GOPs and then storing them in the hierarchical storage of the cloud. For that purpose, the algorithm is carried out at the GOP level of the video stream repository periodically. In the proposed algorithm, we present its pseudo-code. The GOPs of a video stream, GOP transcoding time, GOP size, cloud storage price, and the number of accesses to the video in the last period are received as inputs to the algorithm. The output is to cluster the frequent accessed GOPs/videos, and store them in the cloud storages.

{\SetAlgoNoLine%

\begin{algorithm}

 \SetKwInOut{Input}{Input}
 \SetKwInOut{Output}{Output}

 \Input{Pre-transcoded $GOP_{1} $ to $GOP_{th}$\\
 	   Size of  $GOP_{1} $ to $GOP_{th}$:  $S_{GOP_{j}}$ \\
	   Cloud Storages price: $P_{S_{1}}$,  $P_{S_{2}}$,  $P_{S_{3}}$,  $P_{S_{4}}$\\
	   Number of views of  $ GOP_{1} $ to $GOP_{th}$\\

 }
 \Output{Storage Cost of  pre-transcoded $GOP_{1}$ to $GOP_{th}$\\ }
 
 Apply K-Means clustering on $ GOP_1$ to $GOP_{th}$ with $K=4$\\
 Cluster 1 pre-transcoding cost: $C_{S_{1i}}\leftarrow\dfrac{\sum S_{GOP_{j}} \cdot P_{S_{1}}}{2^{10}}$\\
 Cluster 2 pre-transcoding cost: $C_{S_{2i}}\leftarrow\dfrac{\sum S_{GOP_{j}} \cdot P_{S_{2}}}{2^{10}}$\\
 Cluster 3 pre-transcoding cost: $C_{S_{3i}}\leftarrow\dfrac{\sum S_{GOP_{j}} \cdot P_{S_{3}}}{2^{10}}$\\
 Cluster 4 pre-transcoding cost: $C_{S_{4i}}\leftarrow\dfrac{\sum S_{GOP_{j}} \cdot P_{S_{4}}}{2^{10}}$\\
 Total cost of pre-transcoding $ GOP_1$ to $GOP_{th}$: $C_{S_{GOP_{1}-GOP_{th}}}\leftarrow C_{S_{1i}} + C_{S_{2i}}+ C_{S_{3i}} + C_{S_{4i}} $\\

 \caption{Clustering Pre-transcoding Method}
 \label{al2}
  
\end{algorithm}
}

\begin{figure}
 \includegraphics[width=9cm, scale=1]{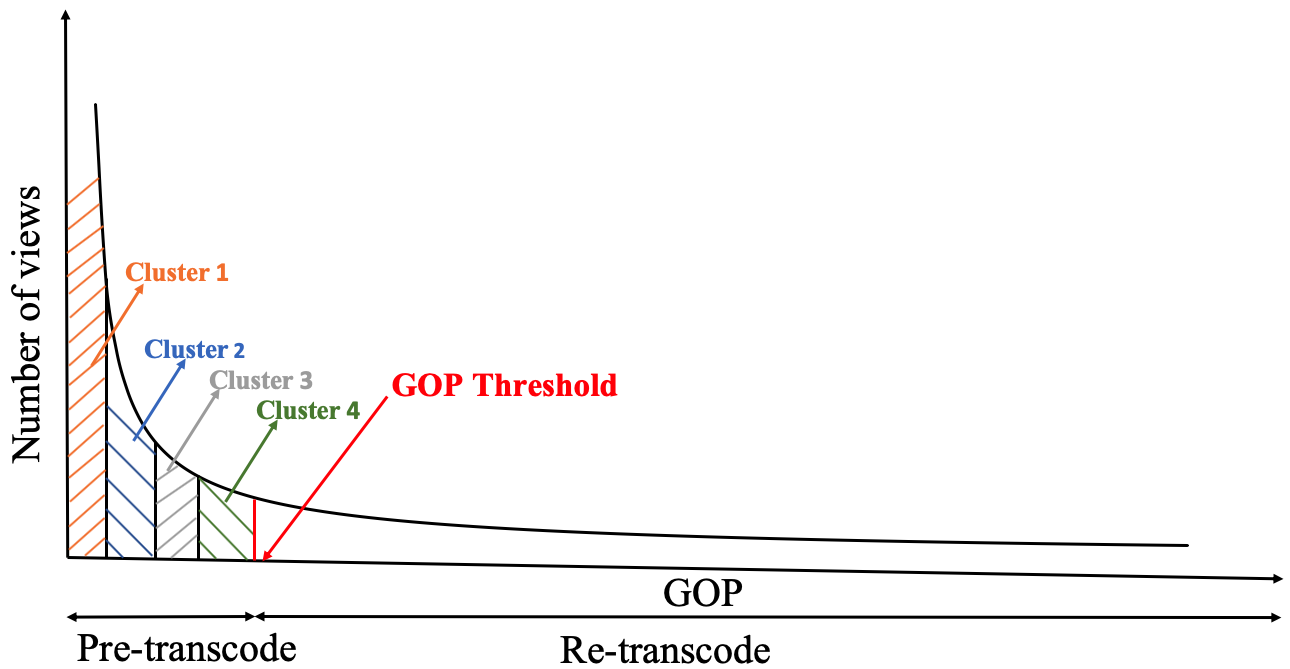}
 \caption{
 Pre-transcoding and Clustering of frequently accessed GOPs in the long-tail distribution}
 \label {Video-partial4}
\end{figure}

The the GOPs' access in a video is a long-tail distribution \cite{miranda} as shown in Fig.\ref{Video-partial4} . The GOPs  before the boundary point ($GOP_{th}$) are pre-transcoded. The algorithm applies the K-means clustering on the pre-transcoded GOPs, it is based on using their number of views as a parameter to decide where each GOP to be stored. The number of clusters of GOPs is 4. In steps (2 - 5), the algorithm calculates the sum of storage cost of GOPs that have similar number of views and stores them in each cluster. In step 6, it sums all the storage costs up of the pre-transcoded GOPs.

\section{Experiment Setup }

\subsection{Videos Synthesis}

Video streams companies use huge repositories to store the videos. We do not have permission to access these repositories. Generally, it is a long and costly process to download a big number of videos and then transcode them. 

Synthesizing videos requires to know their characteristics, particularly, such as GOP size, GOP transcoding time, and number of GOPs for each video. Therefore, we build our repository by executing the method in \cite{darwich2016, xiangbo}.

Based on the obtained characteristics of videos \ie number of GOPs, size of GOP, and the linear equation for GOP transcoding times, we synthesized our repository by generating 50,000 videos.

\subsection{ Amazon Storage Rates}

Amazon offers four types of cloud storage with different price rates and bandwidth accesses as illustrated in table \ref{tab:storageprice}
 \begin{table}[ht]
\centering 
\caption{Amazon storage types and their rates in USD }
\begin{tabular}{c c c c} 
\hline\hline 
 Storage & Price \\ [0.5ex] 
\hline 

\texttt{S3 Standard } & \$0.023 GB/month \\
\texttt{S3 Standard-IA } & \$0.0125 GB/month \\
\texttt{S3 One Zone-IA } & \$0.01 GB/month\\
\texttt{S3 Glacier } & \$0.001 GB/month\\

\hline
\hline 
\end{tabular}

\label{tab:storageprice}
\end{table}

\subsection{ Methods for Comparison}

To assess our proposed method, we use three other methods for comparison. 
\begin {itemize}
\item \emph{Fully pre-transcoding} method, item stores the whole video streams
\item \emph{Fully re-transcoding} method, it re-transcodes all video streams upon request.
\item  Partial pre-transcoding method in \cite{darwich2016} , it partially stores the video stream in the cloud standard storage.

\end{itemize}

\section { Simulation Results}

\subsection{Clustering of FAVs }

\begin{figure}
\centering
 \includegraphics[width=10.0cm]{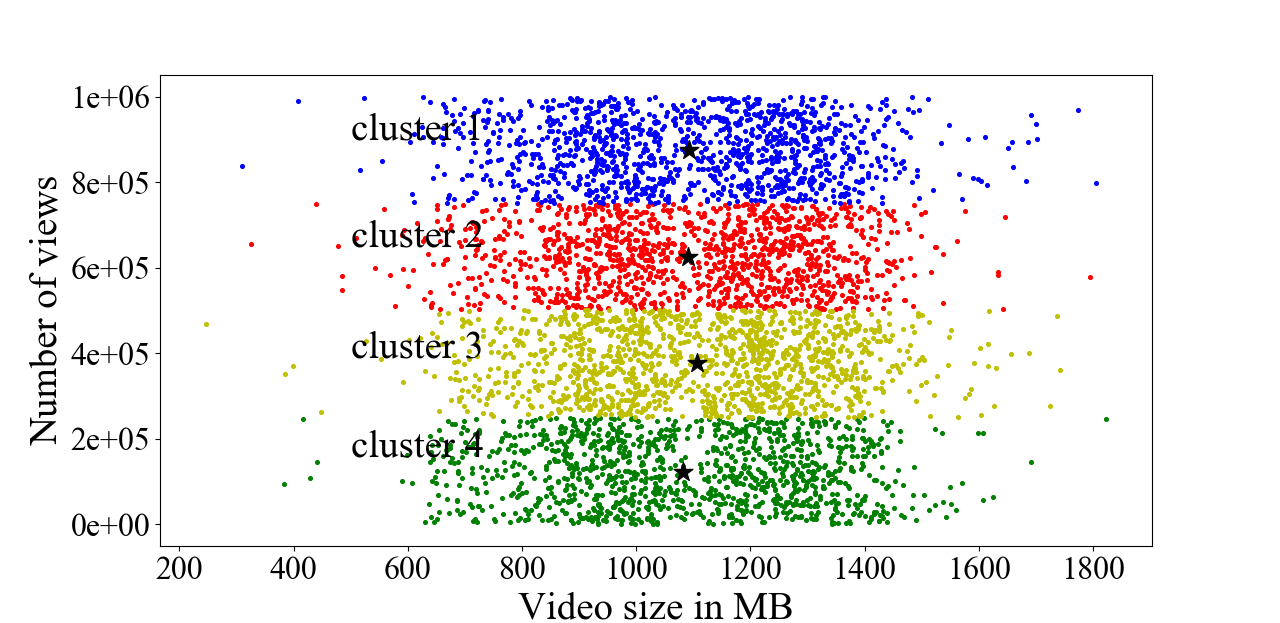}
 \caption{
 Clustering frequently accessed video streams based on the number of views}
 \label {cluster}
\end{figure}

We applied the K-means clustering method on the FAVs in the repository as illustrated in Figure \ref{cluster}. The clustering method groups the frequently accessed videos in four clusters: $Cluster 1$ contains GOPs which have the highest and similar views and uses \texttt{ S3 Standard} to store them, $cluster 2$, $cluster 3$, and $cluster 4$ contains GOPs according to their views similarity and use \texttt{ S3 Standard-IA}, \texttt{S3 One Zone-IA}, and \texttt{S3 glacier} respectively to store these GOPs. the storage types are based on the access bandwidth access and rate. That means the highest number of views for FAVs would be stored in \texttt{ S3 Standard} ($cluster 1$)which has the highest bandwidth access and highest price. Cloud storage \texttt{ S3 Standard-IA}, \texttt{S3 One Zone-IA}, and \texttt{S3 glacier} come subsequently and \texttt{S3 One Zone-IA} ($cluster 4$) has the lowest number of views for FAVs, it provides the lowest bandwidth access and has  lowest price rate.

In this experiment, the x and y-axis represent the video size and number of views respectively. the simulation result showed the GOPs' views of FAVs ranges from $10^3$ to $10^6$. However, the clustering method could be applicable for any different views range of FAVs. Furthermore, we selected the parameter of the clustering $k=4$ because AWS offers 4 types of storage in the cloud.

In our previous work \cite{darwich2016}, we assumed all the pre-transcoded GOPs to be stored in the \texttt{S3 Standard} storage which has the highest price, this incurred a high cost to store all videos in the same storage in the cloud while the proposed clustering method distribute the videos in different cloud storages which cost less.

\subsection{Impact of Changing FAVs number in repository}

To evaluate the proposed method effectively, we need to build a huge repository of videos. Therefore, we synthesize several repositories that contain a total of 50,000 videos each. In such repositories, the percentage of FAVs is varied from 5\% to 30\%.

\begin{figure}
\centering
 \includegraphics[width=9cm]{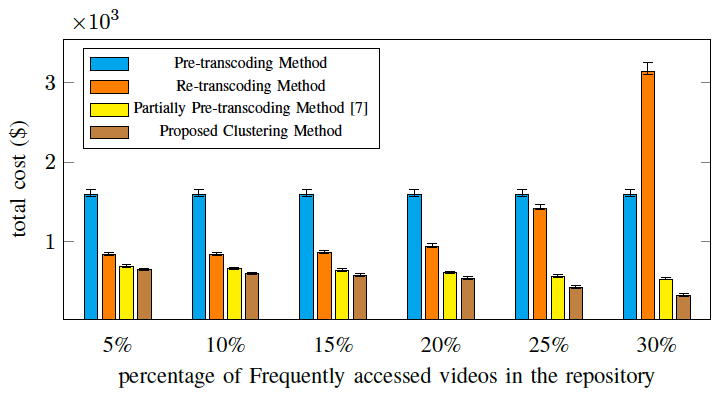}
 \caption{
 Cost comparison of the four methods, full storing method, fully transcoding method, partially pre-transcoding method, and proposed the clustering method when number of frequently accessed video varies}
 \label {favs}
\end{figure}

The simulation result of the total cost of fully pre-transcoding, fully re-transcoding, partial pre-transcoding, and clustering pre-transcoding methods is shown in Fig. \ref{favs}. The method of the full storage does not vary and is constant even the percentage of FAVs changes because the full storage cost does depend on the number of views of videos.

The experimental results in Fig.\ref{favs} show that our proposed clustering method outperforms the other methods and reduces the incurred cost compared to the fully re-transcoding method by up to 90\% when FAVs are 30\% of the repository, also, our proposed method reduces the incurred cost up to 75\% when compared to the fully pre-transcoding method and reduces the cost up to 40\% when compared to the partial pre-transcoding method \cite{darwich2016}. Our proposed method could reduce the cost significantly when the percentage of FAVs increases in the repository.

\section{Conclusion and Future Work}

In this paper, we propose an improved algorithm to minimize further the cost of cloud resources, in particular, we cluster the frequently accessed GOPs/videos and then store them in the four types of cloud storage. We analyze the performance of the proposed when changing the number of FAVs in the repository. Experimental results show the efficiency of our proposed when the number of views of FAVs increases, the incurred cost is reduced up to 40\%.

The future work will be focused on developing video summarization for viewers to improve the quality of services of video streaming.





%


\begin{thebibliography}{1}
\bibitem{report}
G. I. P. Report, “https://www.sandvine.com/trends/global-internetphenomena/,”
accessed October 2019.

\bibitem{forecast}
C. V. N. Index, “Forecast and methodology, 2014-2019,” 2015.
\bibitem{ahmad}
Ahmad, Ishfaq, et al. \emph{``Video transcoding: an overview of various techniques and research issues.''} IEEE Transactions on Multimedia 7.5, 2005.
\bibitem{xiangbo}
Li, Xiangbo, Mohsen Amini Salehi, and Magdy Bayoumi. \emph{``Cloud-based video streaming for energy-and compute-limited thin clients.''} the Stream 2015 Workshop at Indiana University, 2015.
\bibitem{netflix}

http://techblog.netflix.com/2012/12/videos-of-netflix-talks-at-aws-reinvent.html
\bibitem{livlsc}
Li, Xiangbo, Mohsen Amini Salehi, and Magdy Bayoumi. \emph{``VLSC: Video Live Streaming Using Cloud Services.''} 6th IEEE International Conferences on Big Data and Cloud Computing (BDCloud), 2016.
\bibitem{darwich2016}
Darwich, Mahmoud, Ege Beyazit, Mohsen Amini Salehi, and Magdy Bayoumi. \emph{``Cost-efficient repository management for cloud-based on-demand video streaming."} In 2017 5th IEEE International Conference on Mobile Cloud Computing, Services, and Engineering (MobileCloud), pp. 39-44, 2017.


\bibitem{sharma}

N. Sharma, D. K. Krishnappa, D. Irwin, M. Zink, and P. Shenoy \emph{``GreenCache: augmenting off-the-grid cellular towers with multimedia caches.''} Proceedings of the 4th ACM Multimedia Systems Conference, 2013.
\bibitem{li1}
X. Li, M. A. Salehi, M. Bayoumi, and R. Buyya, \emph{``CVSS: A Cost-Efficient and QoS-Aware Video Streaming Using Cloud Services,”} in Proceedings of the 16th ACM/IEEE International Conference on Cluster Cloud and Grid Computing, CCGrid ’16, May 2016.
\bibitem{amazon}
https://aws.amazon.com/ec2/pricing/on-demand/. Accessed October 2019.
\bibitem{jokhio}
F. Jokhio, A. Ashraf, S. Lafond, and J. Lilius, \emph{``A Computation and Storage Trade-off Strategy for Cost-Efficient Video Transcoding in the Cloud,''} 39th Euromicro Conference on Software Engineering and Advanced Applications, 2013.
\bibitem{zhao}
Zhao, Hui, \etal \emph{``A version-aware computation and storage trade-off strategy for multi-version VOD systems in the cloud.''} 20th IEEE Symposium on Computers and Communication (ISCC), 2015.


\bibitem{miranda}
Miranda, Lucas CO, Rodrygo LT Santos, and Alberto HF Laender. \emph{``Characterizing video access s in mainstream media portals.'' } Proceedings of the 22nd International Conference on World Wide Web, 2013.
 
\bibitem{kimmultimedia2019}
Kim, Hyun-Woo, He Mu, Jong Hyuk Park, Arun Kumar Sangaiah, and Young-Sik Jeong. \emph{``Video transcoding scheme of multimedia data-hiding for multiform resources based on intra-cloud."} Journal of Ambient Intelligence and Humanized Computing: pp 1-11, 2019.


\bibitem{darwichhotness}
Mahmoud Darwich, Mohsen Amini Salehi, Ege Beyazit, Magdy Bayoumi, \emph{``Cost-Efficient Cloud-Based Video Streaming Through Measuring Hotness'}, The Computer Journal, Volume 62, Issue 5, pages 641–656, May 2019.

\bibitem{gao}
G. Gao, W. Zhang, Y. Wen, Z. Wang and W. Zhu,\emph{``Towards Cost-Efficient Video Transcoding in Media Cloud: Insights Learned From User Viewings,''} in IEEE Transactions on Multimedia, vol. 17, no. 8, pp. 1286-1296, Aug. 2015.
\bibitem{jokhio1}
Jokhio, Fareed, \etal \emph{``Analysis of video segmentation for spatial resolution reduction video transcoding.''} International Symposium on Intelligent Signal Processing and Communications Systems (ISPACS), 2011.





 

















\end{thebibliography}

\end{document}